\documentclass[10pt]{article}
\usepackage{graphicx}
\graphicspath{{figures/}}
\usepackage{upgreek}
\usepackage{supertabular}
\usepackage{array}
\usepackage{supertabular}
\usepackage{hhline}
\usepackage{graphicx}
\usepackage{xspace}

\newcommand{\vs}{vs.\xspace}

\makeatletter
\newcommand\arraybslash{\let\\\@arraycr}
\makeatother
\usepackage[numbers,sort&compress,round]{natbib}
\usepackage{wrapfig}
\usepackage{siunitx}

\usepackage{amsmath}
\usepackage{amssymb}

\usepackage[labelfont=bf,font=it]{caption}

\usepackage{cleveref}

\usepackage{hyperref}
\hypersetup{%
	colorlinks=true,
	citecolor=blue,
	urlcolor=blue,
  pdfauthor = {Henk W.Ch. Postma}
}

\renewcommand{\and}{;\mbox{ }}
\renewcommand{\textdegree}{$^\circ$}

\usepackage[greek,english]{babel}

\DeclareGraphicsExtensions{.pdf,.jpg}

\title{On the Role of Water Vapor and Process Gasses in Low-Temperature Gold-Catalyzed Graphene Etching}
\author{Ian Carroll, Tanya Klowden, Isabel Alvarez, Henk W.Ch. Postma} 

\usepackage[draft]{fixme}
\usepackage{fixme}
\usepackage{xcolor}

\definecolor{darkgreen}{rgb}{0.0000,0.4000,0.0000}
 \fxsetface{inline}{\color{darkgreen}}
\fxsetface{target}{\color{red}} 
\fxsetface{margin}{\color{darkgreen}} 
\fxsetface{env}{\bfseries} 

\begin{document} 



\newcommand{\fig}{figure }

\twocolumn[
  \begin{@twocolumnfalse}
    \maketitle
    \begin{abstract}
The ability to pattern graphene at low temperatures in a scalable manner is one of the greatest challenges facing graphene industrial adoption today.
We demonstrate a simple method for low-temperature gold-catalyzed etching of graphite with predictable characteristics using ambient air at $350-\SI{375}{\celsius}$. The naturally occurring water vapor in ambient air is necessary for this reaction to occur. In addition, we characterize the etch characteristics as a function of process parameters. Ar annealing is required to obtain crystallographically straight etches, and $\SI{375}{\celsius}$ is required to obtain single-layer deep etching. We anticipate that this work can be adapted by future research to precisely control the shape of the etched areas, allowing for the simple low-temperature creation of nanoscale graphite features, and ultimately can be applied to single-layer graphene sheets for integrated device fabrication. 
\vspace{0.5cm}
    \end{abstract}
 \end{@twocolumnfalse}
]

\section*{Introduction}


The Bergius process has been used for more than a century to hydrogenate coal and produce hydrocarbon fuel \citep{bergius_chemical_1932,bergius_anwendung_1913,bergius_production_1913}. It requires high temperatures and pressure, and the presence of a metal catalyst. It has found other uses as well, from the removal of carbon residue from combustion mechanisms to boosting fuel cell efficiency \citep{zoval_implementation_1996,severin_rapid_2009,lu_highly_2018}.
As the need for nanoscale patterning techniques for graphene has arisen since 2004, various avenues have been taken to investigate this reaction as a potential solution \citep{ci_controlled_2008,campos_anisotropic_2009}. 
The ability of specific metals to provide distinct etch patterns on graphite suggests the potential for catalytic cutting methods to provide a pathway to the high-resolution patterning of graphene that is desired.

Due to graphene's excellent conductivity, tunable electrical properties, and atomic size, there is considerable interest in incorporating it into electronic devices \citep{geim_rise_2007}. Its potential as a replacement for silicon in microprocessors, could ultimately be applied to integrated device fabrication through the simple low-temperature creation of nanoscale graphene features.
Additionally, its high surface area and charge mobility make it a popular candidate for ultracapacitor research  \citep{geim_graphene:_2009}. 
Graphene is a natural choice for future advanced DNA sequencing techniques which rely on its unique thinness and high conductivity to resolve small electrical variations across individual DNA bases passing through a nanogap \citep{postma_rapid_2010}. In order to be realized, all of these ambitions require high-resolution patterning of single-layer graphene.

Previously developed patterning methods include reactive-ion etching \citep{han_energy_2007}, scanning tunneling microscope (STM) lithography \citep{tapaszto_tailoring_2008}, atomic force microscope local oxidation lithography \citep{weng_atomic_2008}, transmission electron microscope drilling \citep{chung_materials_2001,puster_toward_2013}, and chemically-derived techniques \citep{li_chemically_2008}. In general, these techniques are too resource intensive to enable the simple cost-effective production of integrated graphene component devices. High-temperature catalytic etching techniques appear particularly appealing since they can specifically etch in the zig-zag direction \citep{tomita_optical_1974,datta_crystallographic_2008,ci_controlled_2008,campos_anisotropic_2009,datta_wetting_2010}.
However, it requires high temperatures of $900-1100${\textdegree}C for the reaction to take place, limiting its usefulness. Recently, a low-temperature catalytic etching technique was discovered, but the only type of etched shapes that were produced were nanopores and no crystallographic preference was observed \citep{han_scalable_2014,gethers_holey_2015}. Both \citet{han_scalable_2014} and \citet{gethers_holey_2015} argued that oxygen is responsible for Au-catalyzed graphene etching. Here, we demonstrate instead that hydrogen available in naturally present water vapor is responsible. In addition, we also demonstrate that control over the etching characteristics can be attained with careful surface preparation and control of process parameters. This method may ultimately be applied to single-layer graphene sheets for integrated device fabrication.

\section*{Methods and materials}

We tested samples to determine the viability of gold-catalyzed etching, as well as to characterize the reaction and the type of cuts produced.

\subsection*{Surface preparation}

Firstly, we prepared thin flakes of HOPG by separating them from a block of bulk material (NT-MDT Co., ZYB grade, $7 \times 7 \times \SI{1.0}{\milli\meter\cubed}$). The top surface of our HOPG block was ``cleaned'' of amorphous carbon and other contaminants by exfoliating it with blue Nitto tape (Nitto, SPV 224PRM). Next, a thin flake ($0.25-0.5$ mm) of the HOPG was isolated from the bulk mass by shearing it off using a razor blade. This graphite flake was then mounted to a glass microscope slide using double-sided tape and it was placed into a thermal metal evaporator (Thermionics VE-90 Vacuum Evaporator), along with several pellets of $99.9\%$ gold evaporation material (Super Conductor Materials Inc., lot \# 22810-16), and $\SI{2}{\nano\meter}$ of gold were deposited onto the HOPG surface, under a vacuum pressure of $\sim \SI{1e7}{\bar}$ at a rate of $\SI{0.2}{\angstrom\per\second}$. 

\subsection*{Argon anneal}

Each sample was subjected to at least one heat treatment step. It is known that, for catalytic etching to commence, the deposited gold film must be annealed and converted to discrete nanoparticles \citep{datta_crystallographic_2008,ci_controlled_2008,campos_anisotropic_2009,datta_wetting_2010}. In this process, heat is applied to the sample, causing the metal to becoming molten and begin moving across the surface driven by Brownian motion \citep{datta_wetting_2010}. A subset of samples was annealed in a flow of Ar at $\SI{0.600}{\liter\per\second}$. We used a Thermo Scientific Lindberg Blue M furnace, with a maximum temperature of 1100{\textdegree}C, and an MKS Multi Gas Controller Type 647C.

\subsection*{Air etching} 

Next, samples were heated in the presence of ambient air to induce catalysis. The ambient air was temperature controlled to $\SI{21}{\celsius}$ and had a relative humidity level of $20 - 50 \%$ during the months over which these experiments were performed. The ends of our furnace tube were simply disconnected from the gas source, and left open to the room air. Each $25${\textdegree} increment between $25${\textdegree}C and $375${\textdegree}C was tested, to determine the point at which the etching starts. Once we had identified this temperature as $350${\textdegree}C, we tested the effect of temperature and etch duration. 

\subsection*{Gold removal}

Any residual gold was then removed by a $30$ s immersion in KI$_2$-I$_2$ gold etchant (Transene Company Inc., Gold Etch -- Type TFA), followed by two $30$ s immersions in DI water, a $30$ s immersion in isopropyl alcohol, and finally being blown dry.

\subsection*{Control experiments} 

Once we had established the temperature and processing steps to etch the graphite surface, we performed the following control experiments. We demonstrated that gold deposition and Argon anneal only does not cause any etching, showing that it is one of the constituents of air that is responsible. In addition, we tested that an Argon anneal and hot air exposure without Au present does not etch the graphite either, demonstrating that the Au is a necessary component of this etch process. 

\subsection*{Desiccated air}

The effect of the humidity level was tested by exposing the tube surface to desiccated air. Ambient air was pumped through a desiccator and then through a chamber with a hygrometer. The desiccated air was then flushed through the furnace for $\SI{30}{\min}$ while it was held at $\SI{100}{\celsius}$. The furnace was then closed at both ends and raised to the desired `etch' temperature. 

\subsection*{Scan with STM}

\begin{figure}[h]
\includegraphics[width=8cm]{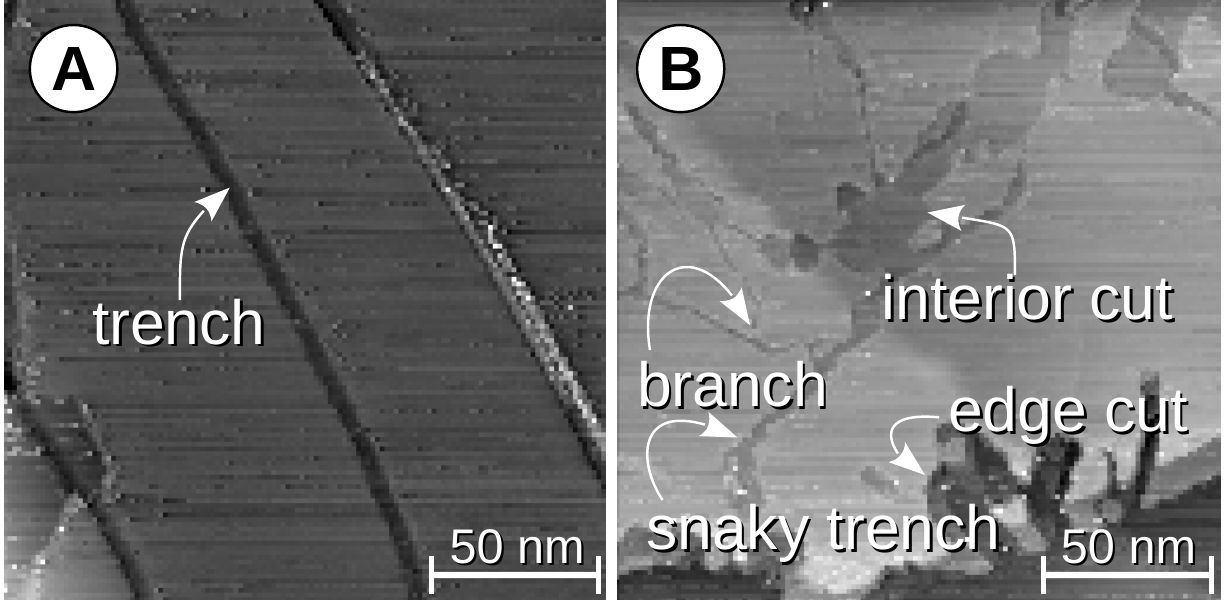}
\caption{\label{fig_stm_image} Representative STM images, obtained by Ar anneal and $\SI{30}{\min}$ etch in air at $\SI{350}{\celsius}$ with etch features as indicated. A) $180\times \SI{180}{\nano\meter\squared}$. B) $230 \times \SI{230}{\nano\meter\squared}$. } 
\end{figure}

All samples were examined with an STM (Nanosurf easyScan 2) for signs of etched areas or other notable changes. The tip voltage was held at $\SI{50}{\milli\volt}$, and the current setpoint was set to $\SI{1}{\nano\ampere}$. These settings were constant across all scans. Images were taken at many scan sizes, from the maximum setting of $1.2 \times \SI{1.2}{\micro\meter\squared}$ down to several $\si{\nano\meter\squared}$. Nanosurf's Easyscan 2 software was used to control the STM and record images [\fig \ref{fig_stm_image}].

\subsection*{Analysis}

Samples which showed significant features as compared to the controls were analyzed using various software packages, including Gwyddion, Adobe Photoshop, and GNU Octave. Etched areas were characterized for their surface area, depth, straightness, and crystallographic orientation. 

\subsubsection*{Surface area etched}

The surface area etched was determined by visually masking off the etched areas using image editing software and converting the number of masked pixels to a physical surface area. 

\begin{figure}[h]
\includegraphics[width=8cm]{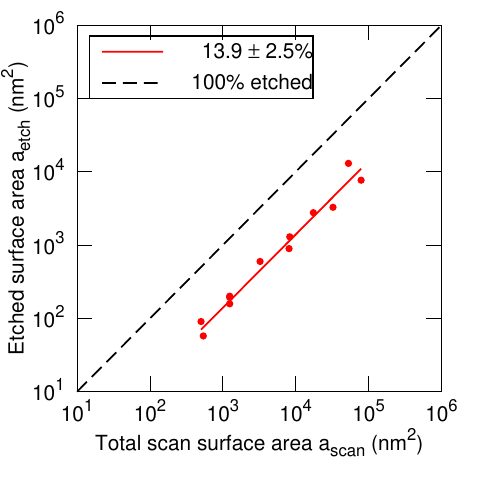}
\caption{\label{area_etch_trend} Linear regressions of area etched to total area scanned with Ar anneal, $\SI{350}{\celsius}$, $\SI{30}{\min}$. The slope of the red trend line gives the percent area etched. Red points mark the proportional etched area of each sample. Dotted line marks a theoretical trend line of $100\%$ area etched.} 
\end{figure}

These quantities were recorded for many scans per individual sample. In order to find a trend correlating the surface area removed with the total scan surface area, a linear regression was performed over all of the data points for each scan [\fig \ref{area_etch_trend}]. The slopes of these regression lines reveals a unique signature for each sample: the percent surface area etched. This quantity was compared across all samples in order to look for trends.

\subsubsection*{Etch depth}

\begin{figure}[h]
\includegraphics[width=7.87cm]{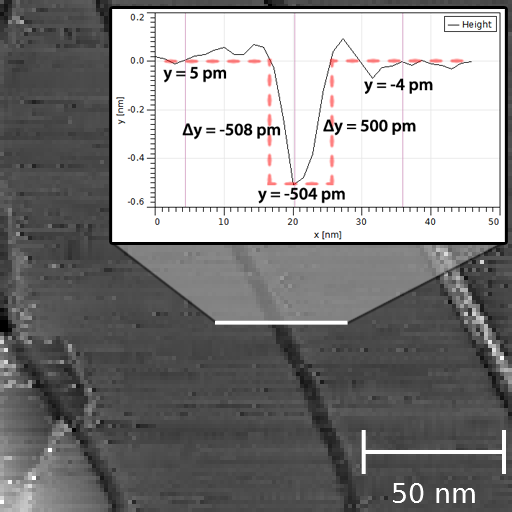}
\caption{\label{fig:etchprofile} STM image of etched trench with inset height profile take across marked white line on image. Black line on height profile indicates measured y-axis height and dotted red line shows calculated y-axis and x-axis measurements adjusted for scanning-tip artifacts. }
\end{figure}

To determine the depth of an etched area, a height profile was taken across it and read at the top and bottom [\fig \ref{fig:etchprofile}]. In the case of trenches, the height difference between both sides of the trench and the bottom were taken and averaged, to account for any slope in the original scan. This depth was then rounded to the nearest graphene step height ($\SI{335}{\pico\meter}$) \citep{burnett_identification_2012} in order to obtain the number of graphene layers removed.

\subsubsection*{Etch direction}

Previous studies of metal-nanoparticle-catalyzed graphene etching have noted a strong crystallographic orientation to the resulting features, with the zig-zag axis being energetically favorable \citep{datta_crystallographic_2008,ci_controlled_2008,campos_anisotropic_2009,datta_wetting_2010}. Seeking to determine if this is also true for gold-catalyzed reactions, we took high-magnification STM scans adjacent to an etched area in order to image the graphene lattice.

\subsubsection*{Etch straightness}

A visual inspection of every etch feature visible was made to determine its degree of straightness. Straight etches are desirable because they indicate strong adherence to a crystallographic axis, and because they are more useful when considering possible uses of this technique. The quantity of straight \vs non-straight etches was calculated as a percentage for each sample, and the percentages for each sample were compared to find any trends correlating percent straight etches and the processing steps used to create that sample.

24 samples were tested across varying processing and heating steps. Of these, only those samples which received a catalytic heating step of $\SI{350}{\celsius}$ or higher showed signs of catalytic etching. Across these samples, $\sim 100$ distinct scans were taken, and $\sim 500$ etched features were examined. 

\section*{Results}

Of the samples which did show etched features, 4 primary types are seen [\fig \ref{fig_stm_image}]; straight etched lines, `trenches', which have a well-defined width and proceed in a constant direction across the length of the scan; `snaky' trenches that show some deviation from straightness, and whose width varies across the scan; branching and amorphous areas, with many small straight lines and large amorphous patches; and `edge cuts', which are rounded areas removed from graphite edges .

\subsection*{Surface area etched}

\begin{figure}[h]
\includegraphics[width=7.87cm]{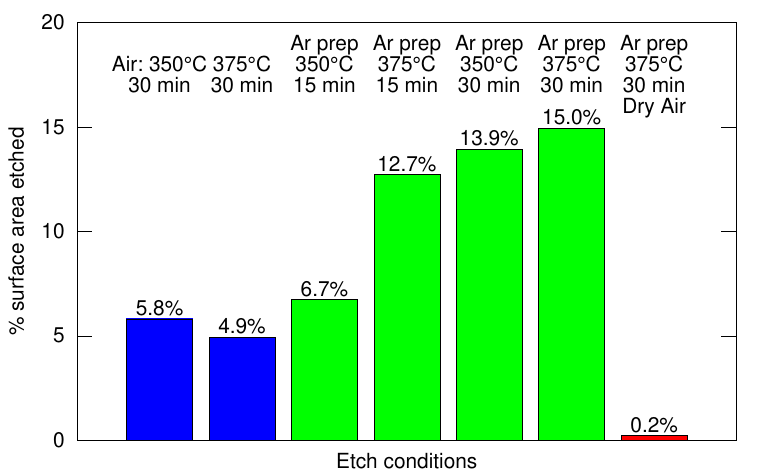}
\caption{\label{fig:area} Etched area for conditions as indicated. Blue and green bars indicated experiments without and with Ar anneal, resp., and atmospheric air conditions. The red bar indicates an Ar anneal and use of dessicated air. }
\end{figure}

In the absence of an Ar anneal, there is no significant difference between etching for $\SI{30}{\min}$ at $350$ and $\SI{375}{\celsius}$ [\fig \ref{fig:area}]. With an Ar anneal step, the amount of surface area etched more than doubles if the etch duration is held at $\SI{30}{\min}$. If the etch duration is reduced to $\SI{15}{\min}$, however, the etched area is reduced for $\SI{350}{\celsius}$. Finally, the amount of surface area etched is almost null when the air is desiccated. 

\subsection*{Etch depth}

\begin{figure}[h]
\includegraphics[width=7.87cm]{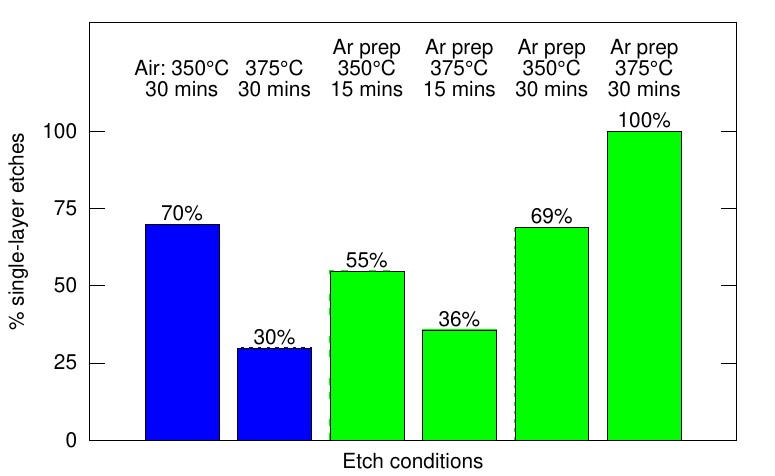}
\caption{\label{fig:singlelayer} Percent single-layer etches per sample for conditions as indicated. Blue and green bars indicated experiments without and with Ar anneal, resp., and atmospheric air conditions. }
\end{figure}

The etch depth is analyzed for all process parameters studied [\fig \ref{fig:singlelayer}]. We find that Ar annealing alone does not cause all etching to be only a single graphene layer deep, while etching at $\SI{375}{\celsius}$ following an Ar anneal causes all etches to be only a single graphene layer deep.

\subsection*{Etch direction}

\begin{figure}[h]
\includegraphics[width=7.87cm]{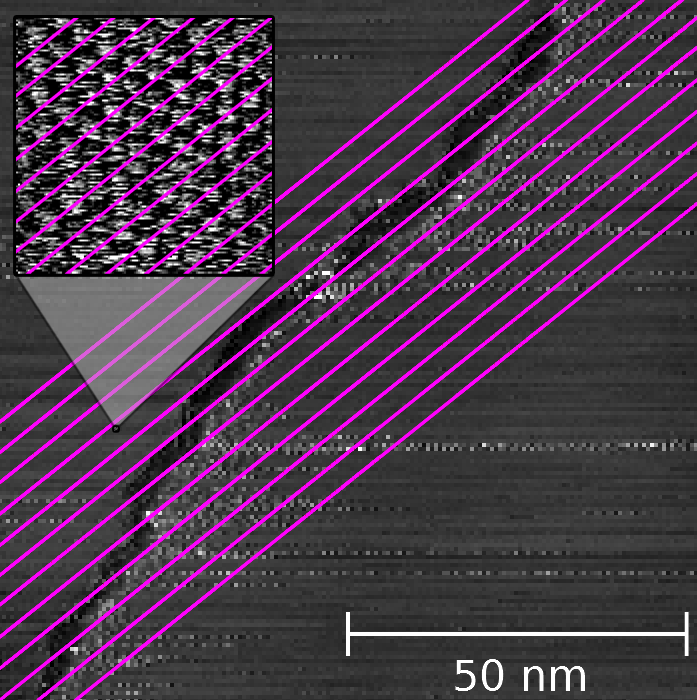}
\caption{\label{fig:etchdirection} STM image of characteristic  zig-zag trench etching. Inset image shows the atomic lattice in the sampled area. Purple bars mark the crystallographic alignment with etching happening preferably along the zig-zag axis.  }
\end{figure}

Atomic resolution images show that etching of trenches occurs predominantly in the zig-zag direction, in line with the work concerning other metal catalyst particles  [\fig \ref{fig:etchdirection}]. 

\subsection*{Etch straightness}

\begin{figure}[h]
\includegraphics[width=7.87cm]{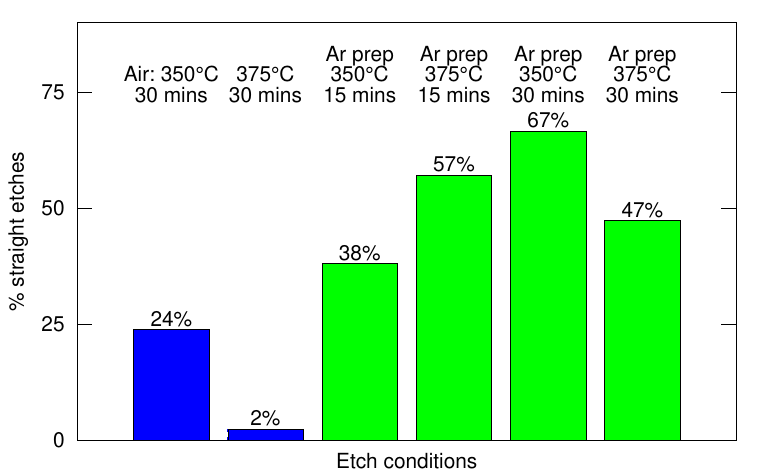}
\caption{\label{fig:straight} Percent straight etches per sample for conditions as indicated. Blue and green bars indicated experiments without and with Ar anneal, resp., and atmospheric air conditions. }
\end{figure}

The percentage of straight etches per sample was determined by visual inspection. Samples that experiences Ar annealing have a higher percentage of straight etches [\fig \ref{fig:straight}].

\subsection*{Ar annealing}

Ar annealing was not necessary for catalytic etching to be performed, although it did have an effect on the quality of the etches. Samples which underwent Ar annealing have larger surface areas etched, and a higher percentage of internal etches and straight etches [\fig \ref{fig:area}, \ref{fig:singlelayer}, \ref{fig:straight}].

\section*{Discussion}

\subsection*{Surface area and catalytic saturation}

We find that the activation energy for the catalytic reaction is not reached until after $\SI{350}{\celsius}$. From there, the etch rate increases with time and temperature; however, the etched amount increases much more dramatically when the time is doubled at $\SI{350}{\celsius}$ than at $\SI{375}{\celsius}$. It is important to note that catalyst particles have a saturation point at which they lose their catalytic ability. 
Beyond this point, they become supersaturated with carbon, and may actually begin to expel carbon nanotubes \citep{zhao_growth_2011,seelan_synthesis_2004}. We conclude that the increase in surface area  etched when going from $15$ to $\SI{30}{\min}$ at $\SI{375}{\celsius}$ is small because by $\SI{15}{\min}$, the catalyst is already approaching its saturation point, and thus the catalytic activity given additional reaction time is limited. At $\SI{350}{\celsius}$, the reaction occurs more slowly, and thus the gains possible with additional time are more significant.

\subsection*{Etch origination and depth}

It appears that the gold catalyst particles are capable of etching downward into the plane of a single graphene layer, but at that point it becomes energetically favorable to continue in a lateral direction. Since they are only consuming carbon from a single layer at a time, these cuts can span large distances. However, cuts that begin at an edge have carbon atoms from all of the layers included in the step height available to them, and thus they can be much deeper. Accordingly, the gold saturates much faster, and these cuts are typically not as long. These results agree with findings with different metal catalysts at higher temperatures \citep{ci_controlled_2008,campos_anisotropic_2009}. 

\subsection*{Ar annealing and catalytic potential}

Annealing in Ar has the effect of increasing the amount etched by a large margin. We argue that the deposited gold film must be annealed into individual droplets before any catalytic cutting can occur. If this happens in the presence of an inert Ar atmosphere, then no carbon is taken up by the gold as it assembles into particles. Once the nanoparticles have formed under Ar, they can begin etching the graphite without already containing carbon, thus having more catalytic potential. If this step is skipped, then the gold must form nanoparticles in a heated air atmostphere, so they begin to pick up carbon even before they form the necessary droplets. Other studies have similarly noted the deleterious effect of carbon-bearing contamination on etch potential \citep{campos_anisotropic_2009}. 

Notably, we observe that among the unannealed samples, more area was etched at $\SI{350}{\celsius}$ than at $\SI{375}{\celsius}$. While this is counter-intuitive in light of previous conclusions, a look at other trends exhibited by the $\SI{375}{\celsius}$ conditions also show the highest percentages of edge etches and multi-layer etches indicating that the catalyst on this sample was mostly spent on etching through the side faces of multi-layer graphene steps.

\subsection*{Crystallographic orientation and straightness}

It appears that the etching of trenches occurs along the zig-zag axis. This is the same observation made by other groups exploring reactions with other metal catalysts \citep{datta_crystallographic_2008,ci_controlled_2008,campos_anisotropic_2009,datta_wetting_2010}. The Ar annealed samples exhibit a higher proportion of straight etches -- ones that are more crystallographically oriented. We posit that annealing the gold film into nanoparticles under Ar prevents them from performing any catalysis until they have achieved a stable, well-defined nanoparticle, which would better hew to a single crystallographic axis when cutting.

\subsection*{Nature of the catalytic reaction}

Both \citet{han_scalable_2014} and \citet{gethers_holey_2015} argued that oxygen is responsible for Au-catalyzed graphene etching that we observe here, but did not provide controls that point to oxygen conclusively. In contrast, we demonstrate naturally present water vapor is responsible for the etching behavior. We assume that the reaction responsible for carbon etching is hydrogenation, and the hydrogen source is chemisorbed on the surface of the gold particles from ambient water vapor, as is seen on nickel surfaces \citep{renouprez_chemisorption_1979}. 

Due to the commonalities shared by these results with those of other metal-nanoparticle catalytic etching studies \citep{datta_crystallographic_2008,ci_controlled_2008,campos_anisotropic_2009,datta_wetting_2010}, it has been assumed here that the reaction responsible for this is the same -- hydrogenation of carbon into methane. 

\section*{Conclusion and Outlook}

Here we have shown that a thin layer of gold, annealed under Ar and heated in the presence of ambient air, can serve as an effective catalyst to produce useful and predictable etch patterns in bulk HOPG at much reduced temperatures compared to previously-investigated metal catalysts. Varying the time, temperature, and pre-processing of this reaction can allow for the results to be tailored for the etch characteristics desired. We demonstrate that water vapor is a necessary ingredient for the etching to occur. Ar annealing is required to obtain crystallographically straight etches, and $\SI{375}{\celsius}$ is required to obtain single-layer deep etching. We believe that this work can be adapted to develop more advanced and specific etching techniques, such as patterning, and can ultimately be adapted to single-layer graphene. This would allow for a simple way to pattern an atomically-thin material at the nanometer scale, while offering flexibility in the fabrication process due to the reduced temperature necessary. Further, by varying the process to favor the supersaturation of the gold nanoparticles instead of single-layer trench etching, we believe that this method could also potentially be adapted to deliberately cultivate carbon nanotube expulsion.

\section*{Figures}

\clearpage 


\begin{thebibliography}{26}
\providecommand{\natexlab}[1]{#1}
\providecommand{\url}[1]{\texttt{#1}}
\expandafter\ifx\csname urlstyle\endcsname\relax
  \providecommand{\doi}[1]{doi: #1}\else
  \providecommand{\doi}{doi: \begingroup \urlstyle{rm}\Url}\fi

\bibitem[Bergius(1932)]{bergius_chemical_1932}
Friedrich Bergius.
\newblock Chemical reactions under high pressure.
\newblock \emph{Nobel Lectures, Chemistry 1922-1941}, pages 244--276, 1932.

\bibitem[Bergius(1913{\natexlab{a}})]{bergius_anwendung_1913}
Friedrich Bergius.
\newblock \emph{Die {Anwendung} hoher drucke bei chemischen {Vorgängen} und
  eine nechbildung des {Entstehungsprozesses} der {Steinkohle}...}
\newblock W. Knapp, 1913{\natexlab{a}}.

\bibitem[Bergius(1913{\natexlab{b}})]{bergius_production_1913}
Friedrich Bergius.
\newblock Production of hydrogen from water and coal from cellulose at high
  temperatures and pressures.
\newblock \emph{Journal of Chemical Technology and Biotechnology}, 32\penalty0
  (9):\penalty0 462--467, 1913{\natexlab{b}}.

\bibitem[Zoval et~al.(1996)Zoval, Biernacki, and
  Penner]{zoval_implementation_1996}
Jim~V. Zoval, Peter~R. Biernacki, and Reginald~M. Penner.
\newblock Implementation of {Electrochemically} {Synthesized} {Silver}
  {Nanocrystallites} for the {Preferential} {SERS} {Enhancement} of {Defect}
  {Modes} on {Thermally} {Etched} {Graphite} {Surfaces}.
\newblock \emph{Analytical Chemistry}, 68\penalty0 (9):\penalty0 1585--1592,
  1996.
\newblock \doi{10.1021/ac951114+}.
\newblock URL \url{https://doi.org/10.1021/ac951114+}.

\bibitem[Severin et~al.(2009)Severin, Kirstein, Sokolov, and
  Rabe]{severin_rapid_2009}
N.~Severin, S.~Kirstein, I.~M. Sokolov, and J.~P. Rabe.
\newblock Rapid {Trench} {Channeling} of {Graphenes} with {Catalytic} {Silver}
  {Nanoparticles}.
\newblock \emph{Nano Letters}, 9\penalty0 (1):\penalty0 457--461, 2009.
\newblock \doi{10.1021/nl8034509}.
\newblock URL \url{https://doi.org/10.1021/nl8034509}.

\bibitem[Lu et~al.(2018)Lu, Xie, Liu, Ai, Zhang, Jin, Zhang, Ma, Li, and
  Shan]{lu_highly_2018}
Peng-Han Lu, De-Gang Xie, Bo-Yu Liu, Fei Ai, Zhao-Rui Zhang, Ming-Shang Jin,
  Xiao~Feng Zhang, Evan Ma, Ju~Li, and Zhi-Wei Shan.
\newblock Highly {Deformable} and {Mobile} {Palladium} {Nanocrystals} as
  {Efficient} {Carbon} {Scavengers}.
\newblock \emph{arXiv:1802.00207 [cond-mat, physics:physics]}, February 2018.
\newblock URL \url{http://arxiv.org/abs/1802.00207}.
\newblock arXiv: 1802.00207.

\bibitem[Ci et~al.(2008)Ci, Xu, Wang, Gao, Ding, Kelly, Yakobson, and
  Ajayan]{ci_controlled_2008}
Lijie Ci, Zhiping Xu, Lili Wang, Wei Gao, Feng Ding, Kevin Kelly, Boris
  Yakobson, and Pulickel Ajayan.
\newblock Controlled nanocutting of graphene.
\newblock \emph{Nano Research}, 1\penalty0 (2):\penalty0 116--122, 2008.
\newblock \doi{10.1007/s12274-008-8020-9}.
\newblock URL \url{http://dx.doi.org/10.1007/s12274-008-8020-9}.

\bibitem[Campos et~al.(2009)Campos, Manfrinato, Sanchez-Yamagishi, Kong, and
  Jarillo-Herrero]{campos_anisotropic_2009}
Leonardo~C. Campos, Vitor~R. Manfrinato, Javier~D. Sanchez-Yamagishi, Jing
  Kong, and Pablo Jarillo-Herrero.
\newblock Anisotropic {Etching} and {Nanoribbon} {Formation} in
  {Single}-{Layer} {Graphene}.
\newblock \emph{Nano Letters}, 9\penalty0 (7):\penalty0 2600--2604, July 2009.
\newblock \doi{10.1021/nl900811r}.
\newblock URL \url{http://dx.doi.org/10.1021/nl900811r}.

\bibitem[Geim and Novoselov(2007)]{geim_rise_2007}
A.~K. Geim and K.~S. Novoselov.
\newblock The rise of graphene.
\newblock \emph{Nature materials}, 6\penalty0 (3):\penalty0 183--191, 2007.

\bibitem[Geim(2009)]{geim_graphene:_2009}
A.~K. Geim.
\newblock Graphene: status and prospects.
\newblock \emph{Science}, 324\penalty0 (5934):\penalty0 1530--1534, 2009.

\bibitem[Postma(2010)]{postma_rapid_2010}
Henk W.~Ch. Postma.
\newblock Rapid {Sequencing} of {Individual} {DNA} {Molecules} in {Graphene}
  {Nanogaps}.
\newblock \emph{Nano Letters}, 10\penalty0 (2):\penalty0 420--425, February
  2010.
\newblock \doi{10.1021/nl9029237}.
\newblock URL \url{http://dx.doi.org/10.1021/nl9029237}.

\bibitem[Han et~al.(2007)Han, Ozyilmaz, Zhang, and Kim]{han_energy_2007}
Melinda~Y. Han, Barbaros Ozyilmaz, Yuanbo Zhang, and Philip Kim.
\newblock Energy {Band}-{Gap} {Engineering} of {Graphene} {Nanoribbons}.
\newblock \emph{Physical Review Letters}, 98\penalty0 (20):\penalty0 206805--4,
  May 2007.
\newblock URL \url{http://link.aps.org/abstract/PRL/v98/e206805}.

\bibitem[Tapaszto et~al.(2008)Tapaszto, Dobrik, Lambin, and
  Biro]{tapaszto_tailoring_2008}
Levente Tapaszto, Gergely Dobrik, Philippe Lambin, and Laszlo~P. Biro.
\newblock Tailoring the atomic structure of graphene nanoribbons by scanning
  tunnelling microscope lithography.
\newblock \emph{Nat Nano}, 3\penalty0 (7):\penalty0 397--401, July 2008.
\newblock ISSN 1748-3387.
\newblock \doi{10.1038/nnano.2008.149}.
\newblock URL \url{http://dx.doi.org/10.1038/nnano.2008.149}.

\bibitem[Weng et~al.(2008)Weng, Zhang, Chen, and Rokhinson]{weng_atomic_2008}
Lishan Weng, Liyuan Zhang, Yong~P. Chen, and L.~P. Rokhinson.
\newblock Atomic force microscope local oxidation nanolithography of graphene.
\newblock \emph{Applied Physics Letters}, 93\penalty0 (9):\penalty0 093107--3,
  2008.
\newblock \doi{10.1063/1.2976429}.
\newblock URL \url{http://link.aip.org/link/?APL/93/093107/1}.

\bibitem[Chung(2001)]{chung_materials_2001}
D.~D.~L. Chung.
\newblock Materials for thermal conduction.
\newblock \emph{Applied thermal engineering}, 21\penalty0 (16):\penalty0
  1593--1605, 2001.
\newblock URL
  \url{http://www.sciencedirect.com/science/article/pii/S1359431101000424}.

\bibitem[Puster et~al.(2013)Puster, Rodríguez-Manzo, Balan, and
  Drndic]{puster_toward_2013}
Matthew Puster, Julio~A. Rodríguez-Manzo, Adrian Balan, and Marija Drndic.
\newblock Toward sensitive graphene nanoribbon–nanopore devices by preventing
  electron beam-induced damage.
\newblock \emph{ACS nano}, 7\penalty0 (12):\penalty0 11283--11289, 2013.
\newblock URL \url{http://pubs.acs.org/doi/abs/10.1021/nn405112m}.

\bibitem[Li et~al.(2008)Li, Wang, Zhang, Lee, and Dai]{li_chemically_2008}
Xiaolin Li, Xinran Wang, Li~Zhang, Sangwon Lee, and Hongjie Dai.
\newblock Chemically {Derived}, {Ultrasmooth} {Graphene} {Nanoribbon}
  {Semiconductors}.
\newblock \emph{Science}, 319\penalty0 (5867):\penalty0 1229--1232, February
  2008.
\newblock \doi{10.1126/science.1150878}.
\newblock URL
  \url{http://www.sciencemag.org/cgi/content/abstract/319/5867/1229}.

\bibitem[Tomita and Tamai(1974)]{tomita_optical_1974}
Akira Tomita and Yasukatsu Tamai.
\newblock Optical microscopic study on the catalytic hydrogenation of graphite.
\newblock \emph{The Journal of Physical Chemistry}, 78\penalty0 (22):\penalty0
  2254--2258, 1974.

\bibitem[Datta et~al.(2008)Datta, Strachan, Khamis, and
  Johnson]{datta_crystallographic_2008}
Sujit~S. Datta, Douglas~R. Strachan, Samuel~M. Khamis, and A.~T.~Charlie
  Johnson.
\newblock Crystallographic {Etching} of {Few}-{Layer} {Graphene}.
\newblock \emph{Nano Letters}, 8\penalty0 (7):\penalty0 1912--1915, July 2008.
\newblock \doi{10.1021/nl080583r}.
\newblock URL \url{http://dx.doi.org/10.1021/nl080583r}.

\bibitem[Datta(2010)]{datta_wetting_2010}
Sujit~S. Datta.
\newblock Wetting and energetics in nanoparticle etching of graphene.
\newblock \emph{Journal of Applied Physics}, 108\penalty0 (2):\penalty0 024307,
  2010.
\newblock ISSN 00218979.
\newblock \doi{10.1063/1.3456100}.
\newblock URL \url{http://link.aip.org/link/JAPIAU/v108/i2/p024307/s1&Agg=doi}.

\bibitem[Han et~al.(2014)Han, Funk, Shen, Chen, Li, Campbell, Dai, Yang, Kim,
  Liao, Connell, Barone, Chen, Lin, and Hu]{han_scalable_2014}
Xiaogang Han, Michael~R. Funk, Fei Shen, Yu-Chen Chen, Yuanyuan Li, Caroline~J.
  Campbell, Jiaqi Dai, Xiaofeng Yang, Jae-Woo Kim, Yunlong Liao, John~W.
  Connell, Veronica Barone, Zhongfang Chen, Yi~Lin, and Liangbing Hu.
\newblock Scalable {Holey} {Graphene} {Synthesis} and {Dense} {Electrode}
  {Fabrication} toward {High}-{Performance} {Ultracapacitors}.
\newblock \emph{ACS Nano}, 8\penalty0 (8):\penalty0 8255--8265, August 2014.
\newblock ISSN 1936-0851.
\newblock \doi{10.1021/nn502635y}.
\newblock URL \url{http://dx.doi.org/10.1021/nn502635y}.

\bibitem[Gethers et~al.(2015)Gethers, Thomas, Jiang, Weiss, Duan, Goddard, and
  Weiss]{gethers_holey_2015}
Matthew~L. Gethers, John~C. Thomas, Shan Jiang, Nathan~O. Weiss, Xiangfang
  Duan, William~A. Goddard, and Paul~S. Weiss.
\newblock Holey {Graphene} as a {Weed} {Barrier} for {Molecules}.
\newblock \emph{ACS Nano}, 9\penalty0 (11):\penalty0 10909--10915, November
  2015.
\newblock ISSN 1936-0851.
\newblock \doi{10.1021/acsnano.5b03936}.
\newblock URL \url{http://dx.doi.org/10.1021/acsnano.5b03936}.

\bibitem[Burnett et~al.(2012)Burnett, Yakimova, and
  Kazakova]{burnett_identification_2012}
Tim~L. Burnett, Rositza Yakimova, and Olga Kazakova.
\newblock Identification of epitaxial graphene domains and adsorbed species in
  ambient conditions using quantified topography measurements.
\newblock \emph{Journal of Applied Physics}, 112\penalty0 (5):\penalty0 054308,
  2012.
\newblock URL \url{http://aip.scitation.org/doi/abs/10.1063/1.4748957}.

\bibitem[Zhao and He(2011)]{zhao_growth_2011}
Yanli Zhao and Suxiang He.
\newblock The growth of carbon nanotube with chemical vapor deposition under
  different process parameters.
\newblock \emph{Journal of Wuhan University of Technology-Mater. Sci. Ed.},
  26\penalty0 (2):\penalty0 202, 2011.
\newblock URL \url{http://link.springer.com/article/10.1007/s11595-011-0197-1}.

\bibitem[Seelan et~al.(2004)Seelan, Hwang, Hwang, and
  Sinha]{seelan_synthesis_2004}
S.~Seelan, D.~W. Hwang, L.-P. Hwang, and A.~K. Sinha.
\newblock Synthesis of multiwalled carbon nanotubes by high-temperature vacuum
  annealing of amorphous carbon.
\newblock \emph{Vacuum}, 75\penalty0 (2):\penalty0 105--109, 2004.
\newblock URL
  \url{http://www.sciencedirect.com/science/article/pii/S0042207X04001253}.

\bibitem[Renouprez et~al.(1979)Renouprez, Fouilloux, Candy, and
  Tomkinson]{renouprez_chemisorption_1979}
A.~J. Renouprez, P.~Fouilloux, J.~P. Candy, and J.~Tomkinson.
\newblock Chemisorption of water on nickel surfaces.
\newblock \emph{Surface Science}, 83\penalty0 (1):\penalty0 285--295, 1979.
\newblock URL
  \url{http://www.sciencedirect.com/science/article/pii/003960287990493X}.

\end{thebibliography}
\end{document}